\documentclass[prd,aps]{revtex4}
\newcommand{\ba}{\begin{eqnarray}}
\newcommand{\ea}{\end{eqnarray}}
\newcommand{\by}{\begin{array}}
\newcommand{\ey}{\end{array}}
\newcommand{\f}{\frac}
\newcommand{\p}{\partial}
\newcommand{\om}{\omega}
\newcommand{\s}{\sqrt}
\newcommand{\iy}{\infty}

\begin{document}
\title{Weakly bound electrons in external magnetic field}
\author{I.V.Mamsurov$^1$ and F. Kh. Chibirova$^2$}
\affiliation{$^1$Faculty of Physics, Moscow State University,
119899, Moscow, Russia\\ $^2$Karpov Institute of Physical
Chemistry, 103064, Moscow, Russia}

\begin{abstract}
The effect of the uniform
magnetic field on the electron in the
spherically symmetric square-well  potential is studied.
A transcendental equation that determines the electron
energy spectrum is derived. The approximate value
of the lowest (bound) energy state is found.
The approximate wave function and probability
current density of this state are constructed.
\end{abstract}


\maketitle

\section{Introduction}
Quantum nonrelativistic systems in external electromagnetic field
have attracted permanent interest due to possible application of
their models in many phenomena of quantum mechanics. Particularly
it concerns so-called bound electron states. For example, it is
well known that the integer quantum Hall effect is correlated with
the presence of weakly bound electron states in corresponding
samples. Nonrelativistic electrons in an external magnetic field
are also responsible for such remarkable macroscopic quantum
phenomena as, for example, high-temperature superconductivity.
\cite{W}. Magnetic fields are also likely to effect on weakly
bound electrons  into singular potentials of defects in defect
films \cite{Ch1,Ch2}. The effect of magnetic fields on loosely
bound electron in two dimensions models was studied in \cite{Kh}.
To this problem are also related such phenomena as parity
violation, the Aharonov-Bohm effect \cite{Ah}, and others. The
behavior of an electron in a constant uniform magnetic field and
single attractive $\delta$ potential in three spatial dimensions
was studied in \cite {Kh1}, in which it was also obtained non
trivial result for probability current density of the loosely
bound electron state. This current resembles "pancake vortices" in
the high-temperature superconductors.

In this paper is studied a more general case of the electron
behavior in external uniform magnetic field in the presence of
spherically symmetric square-well  potential of finite radius. The
calculations are made supposing small size of this radius compared
to the magnetic length $a=\s{\hbar/m \om}$. In the first
approximation in the small parameter $\xi =R^2/2a^2\ll 1$ is
derived transcendental equation for the electron energy spectrum,
and also approximate value of the bound energy state. Accordingly
in zero approximation a wave function of the bound state is
obtained, and the value of probability current for this state is
calculated. This current, as also in  \cite{Kh1}, appears to have
non zero circulation around the axis parallel to the external
magnetic field and to be mostly confined within the perpendicular
plane.

\section{Schr\"odinger-Pauli Equation }
Let us consider an electron in a spherically symmetric square-well
potential of the form: \ba U(r)= \left\{ \by {c} -U_0,r<R \\ 0,r>R
\ey \right. \ea in the presence of uniform magnetic field $H$,
which is directed along the axis $z$. Vector potential is
specified in a cylindrically symmetric gauge: \ba
A_{\phi}=\frac{H\rho}{2}, A_{\rho}=A_{z}=0. \ea Let us write the
Schr\"odinger-Pauli equation for this electron: \ba
i\hbar\frac{\partial}{\partial t}\psi(t, {\bf r})=\hat{H} \psi(t,
{\bf r}), \label{Pauli} \ea where Hamiltonian in cylindrical
coordinates has the form: \ba \hat{H}=-\frac{\hbar^2}{2m} \left[
\f{\p}{\rho \p \rho} \left( \f{\rho \p}{\p \rho} \right)
+\f{\p^2}{\p z^2} + \f{\p^2}{\rho^2 \p \phi^2} \right] -\f{i\hbar
\om}{2} \f{\p}{\p \phi} +\f{m\om^2}{8} \rho^2 +U(\s{\rho^2+z^2})
+\mu \sigma_3 H, \label{Ham} \ea where: \ba \om=\f{|e|m}{Hc},
\mu=\f{|e|\hbar}{2mc}, \sigma_3= \left( \by {c}
1 \ 0\\
0 \ 1\\
\ey \right). \ea We are interested in a stationary solution of the
equation (\ref{Pauli}): \ba \psi(t, {\bf r})=e^{\f{-iEt}{\hbar}}
\psi_E ({\bf r}). \label{wf} \ea It is reasonable to seek the
spatial part of the wave function in the form: \ba \psi_E ({\bf
r}) =\int\limits_{-\iy}^{+\iy} dp_z \sum_{l=-\iy}^{+\iy}
\sum_{n_{\rho}=0}^{\iy} C_{En_{\rho} l p_z} \psi_{n_{\rho} l p_z}
({\bf r}). \label{repr} \ea Wave functions on the right side of
this equation (\ref{repr}) are eigenfunctions of Hamiltonian
(\ref{Ham}) in the absence of spherically symmetric potential
(see, for example, \cite{L}). \ba \psi_{n_{\rho} l p_z} ({\bf
r})=\f{1}{2} \f{e^{i p_z z/ \hbar}}{\s{2\pi\hbar}}
 \f{e^{il\phi}}{\s{2\pi}} \f{1}{a} I_{n_{\rho} l} (\rho^2 /2a^2)
\left( \by {c}
1+s \\
1-s \\
\ey \right). \ea Where $s=\pm 1$ is a constant quantum spin number
of an electron, $a=\s{\hbar/m\om}$, and Laguerre functions: \ba
I_{n_{\rho} l}  (x) =\f{1}{\s{(n_{\rho} + |l|)! n_{\rho}!}}
e^{-x/2} x^{|l|/2} Q^{|l|}_{n_{\rho}}(x) \ea are expressed through
corresponding polynomials. Multiplying the both sides of
(\ref{repr}) by $\psi_{N_{\rho} L P_z}$, transferring to the right
side the term containing the potential $U({\bf r})$, and to the
left side all other terms, and integrating  over all spatial
coordinates, we obtain: \ba C_{E N_{\rho} L P_z} \left( \hbar \om
\left( N_{\rho} +
\f{|L|+L+1+s}{2}\right) +\f{P^2_z}{2m} - E \right)= \nonumber \\
=\f{U_0}{\pi} \int\limits^{+\iy}_{-\iy} dp_z \sum^{\iy}_{n_{\rho}=0}
C_{E n_{\rho} L p_z} \f{1}{P_z -p_z} \nonumber \\
\f{1}{a^2} \int\limits^{R}_{0} \rho d\rho I_{n_{\rho} L}
(\rho^2/2a^2) I_{N_{\rho} L} (\rho^2/2a^2) \sin \left(
\f{P_z-p_z}{\hbar} \s{R^2 - \rho^2} \right). \label{Eq} \ea Taking
into account a small radius of the potential well compared with
magnetic length: $R^2/a^2<<1$, we expand the product of Laguerre
functions on the right side of (\ref{Eq}) in a power series of
parameter $\rho^2/2a^2$, up to the terms of the first order. We
also suppose that in reality the integral over $p_z$ on the right
side of (\ref{Eq}) has the finite limits. Accordingly we put:
$\s{R^2-\rho^2} (P_z-p_z)/\hbar<<1$ and substitute the sinus on
the right side of (\ref{Eq}) by its argument. Then integrating
over $\rho$ and using designation: $U_0 R^3=\lambda$, we obtain
the following result: \ba C_{E N_{\rho} 0 P_z} \left( \hbar \om
\left( N_{\rho} +
\f{1+s}{2}\right) +\f{P^2_z}{2m} - E \right) = \nonumber \\
=\f{\lambda}{\pi \hbar a^2} \int\limits^{+\iy}_{-\iy} dp_z \sum^{\iy}_{n_{\rho}=0}
C_{E n_{\rho} 0 p_z}
\left[ \f{1}{3}
- \f{2}{15} \xi (1+n_{\rho} + N_{\rho}) \right], L=0.
\label{Eq1}
\ea

\ba
C_{E N_{\rho} L P_z}
\left( \hbar \om \left( N_{\rho} +
\f{|L|+L+1+s}{2}\right) +\f{P^2_z}{2m} - E \right) = \nonumber \\
=\f{\lambda}{\pi \hbar a^2} \int\limits^{+\iy}_{-\iy} dp_z \sum^{\iy}_{n_{\rho}=0}
C_{E n_{\rho} L p_z}
\left[
\f{2}{15} \xi \s{(n_{\rho} +1) (N_{\rho} + 1)} \right], L=\pm1
\label{Eq2}
\ea
Where $\xi =m \om R^2 /2\hbar$.
In case of all other values of  $L$ the right side of  (\ref{Eq})
in the first approximation equals to zero.
It means that corresponding coefficients $C_{E n_{\rho} L p_z}$
also equal to zero when $L \ne 0, \pm 1$.
\section{Energy Spectrum}
Because we are interested in the lowest energy state, we consider
only the case $L=0$. We seek coefficients $C_{E n_{\rho} 0 p_z}$
in the form: \ba C_{E n_{\rho} 0 p_z}= C_E \f{1- \f{2}{5} \xi (1/2
+ n_{\rho})}{\hbar \om \left( n_{\rho} + \f{1+s}{2}\right)
+\f{p^2_z}{2m} - E}, \label{Coef} \ea Inserting (\ref{Coef}) in
(\ref{Eq1}) and neglecting the term proportional to $\xi^2$, we
obtain the equation for energy spectrum: \ba 1 = \f{\lambda}{3\pi
\hbar a^2} \int\limits^{+\iy}_{-\iy} dp_z \sum^{\iy}_{n_{\rho}=0}
\f{1 - \f{4}{5} \xi (1/2 + n_{\rho})}{\hbar \om \left( n_{\rho} +
\f{1+s}{2}\right) +\f{p^2_z}{2m} - E}. \label{Sp} \ea Integrating
over $p_z$ we finally have: \ba 1=\f{\s{2m}}{3 \hbar a^2} \lambda
\sum^{\iy}_{n_{\rho}=0} \f{1- \f{4}{5} \xi (1/2
+n_{\rho})}{\s{\hbar \om \left( n_{\rho} + \f{1+s}{2}\right) -
E}}. \label{Sp1} \ea

This equation may be
solved graphically.
In order to find the approximate value
of lowest (bound) state, we put
in (\ref{Sp1}): $n_{\rho}=0, s=-1$. Then we obtain
the following result:
\ba
E_{min}=-\f{2m\lambda^2}{9 \hbar^2 a^4} \left( 1-\f{2}{5} \xi \right).
\label{Lev}
\ea
\section{Wave Function and Probability Current}
Let us find in zero approximation of $\xi$ the wave function
$\psi_E ({\bf r})$ for lowest energy state. In this case in
expansion of the product of Laguerre functions in (\ref{Eq}) we
consider only one term, which does not contain $\xi$. Then the
right side of (\ref{Eq}) does not equal to zero only when $L=0$.
In the formula for coefficients $C_{n_{\rho} 0 p_z}$ (\ref{Coef})
we must neglect the member proportional to $\xi$. We find
coefficient  $C_E$ from the normalizing equation: \ba
\int\limits_{-\iy}^{+\iy} dp_z \sum_{l=-\iy}^{+\iy}
\sum_{n_{\rho}=0}^{\iy} |C_{En_{\rho}l p_z}|^2 =1. \label{Norm}
\ea Taking into account that in the summation over $l$ only one
term of zero order is present, after integration over  $p_z$ we
have: \ba C_E=\f{1}{m \s{2\pi}} \left[ \sum^{\iy}_{n_{\rho}=0}
\f{1}{(\hbar \om n_{\rho} -E)^{3/2}} \right]^{-1/2}. \label{C_E}
\ea Inserting (\ref{C_E}) in (\ref{Coef}), and (\ref{Coef}) in
(\ref{repr}), and again taking into account that in the summation
over $l$ only term of zero order is rest, we obtain the following
formula: \ba \psi_{E, s=-1}=\f{1}{2\pi a \s{\hbar}} C_E
\sum^{\iy}_{n_{\rho}=0} \int\limits^{+\iy}_{-\iy} dp_z \f{1}{\hbar
\om n_{\rho} +\f{p^2_z}{2m} - E} e^{\f{i p_z z}{\hbar}}
I_{n_{\rho} 0} (\rho^2/2a^2). \label{F} \ea Because we are
interested in the lowest (bound) state, we consider only term with
$n_{\rho}=0$. Then integrating over $p_z$ we finally obtain: \ba
\psi_{E, s=-1}=\f{1}{2\pi a \s{2m \hbar |E|} } C_E
\exp\left(-\s{2m|E|} \f{\theta (z) z}{\hbar}\right)
\exp\left(-\f{m \om \rho^2}{4\hbar}\right). \label{F1} \ea Where
$\theta (z) = 1 (-1)$ when $z>0 (<0)$. Using well known expression
for the density of probability current: \ba {\bf j} =\f{i
\hbar}{2m} (\psi \nabla \psi^* -\psi^* \nabla\psi) - \f{e}{mc}
{\bf A} \psi^* \psi, \label{Cur} \ea we obtain the following
result: \ba j_{\phi}=-\f{eH\rho}{16\pi^2 a^2 m^2 c \hbar  |E|}
C_E^2 \exp\left(-2\s{2m|E|} \f{\theta (z) z}{\hbar}\right)
\exp\left(-\f{m \om \rho^2}{2\hbar}\right),j_{\rho}=j_z=0.
\label{Cur1} \ea
\section{Discussion}
So it is established that the presence of external magnetic field
and potential well of finite depth produces an interesting  bound
energy state of an electron. Its probability current has nonzero
circulation along the field axis. This fact  attracts significant
interest for it contributes to explanation of many quantum
mechanics phenomena, in the first place such as high-temperature
superconductivity.

\vspace{0.5cm} \noindent{\bf Acknowledgments}

This paper was supported  by a Joint Research Project  of the
Taiwan National Science Council (NSC-RFBR No. 95WFD0400022,
Republic of China) and the Russian Foundation for Basic Research
(No. NSC-a-89500.2006.2) under contract No. RP06N04-1, by  the
U.S. Department of Energy's Initiative for Proliferation
Prevention (IPP) Program through Contract No. 94138 with the
Brookhaven National Laboratory, and, in part, by the Program for
Leading Russian Scientific Schools (Grant No.
NSh-5332.2006.2)(I.V. M.).

The authors are grateful to V. R. Khalilov for fruitful discussions.

\end{document}